\documentclass[]{spie}  
\usepackage[]{graphicx}

\title{Design and tests of the hard X-ray polarimeter X-Calibur}

\author{M.~Beilicke\supit{a},
M.G.~Baring\supit{b},
S.~Barthelmy\supit{c},
W.R.~Binns\supit{a},
J.~Buckley\supit{a},
R.~Cowsik\supit{a},
P.~Dowkontt\supit{a},
A.~Garson\supit{a},
Q.~Guo\supit{a},
Y.~Haba\supit{d},
M.H.~Israel\supit{a},
H.~Kunieda\supit{d},
K.~Lee\supit{a},
H.~Matsumoto\supit{d},
T.~Miyazawa\supit{d},
T.~Okajima\supit{c},
J.~Schnittman\supit{c},
K.~Tamura\supit{d},
J.~Tueller\supit{c},
and H.~Krawczynski\supit{a}
\skiplinehalf
\supit{a}Department of Physics and McDonnell Center for the Space Sciences, Washington University, St. Louis, MO, USA;
\supit{b}Rice University, TX, USA;
\supit{c}Goddard Space Flight Center, MD, USA;\\
\supit{d}Nagoya University, Japan;
}

\authorinfo{Further author information: (Send correspondence to
M.\,Beilicke): beilicke@physics.wustl.edu}

\begin{document} 
  \maketitle 

\begin{abstract}

X-ray polarimetry promises to give qualitatively new information about
high-energy astrophysical sources, such as binary black hole systems,
micro-quasars, active galactic nuclei, and gamma-ray bursts. We designed,
built and tested a hard X-ray polarimeter {\it X-Calibur} to be used in
the focal plane of the InFOC$\mu$S grazing incidence hard X-ray
telescope. X-Calibur combines a low-Z Compton scatterer with a CZT
detector assembly to measure the polarization of $10-80 \, \rm{keV}$
X-rays making use of the fact that polarized photons Compton scatter
preferentially perpendicular to the electric field orientation.
X-Calibur achieves a high detection efficiency of order unity.

\end{abstract}



\section{INTRODUCTION}
\label{sec:intro}

\paragraph{Motivation.} Spectral and morphological studies in the X-ray
energy band (and above) have become established tools to study the
non-thermal emission processes of various astrophysical sources.
However, many of the regions of interest (black hole vicinities, jet
formation zones, etc.) are too small to be spatially resolved with
current and future instruments. Spectropolarimetric X-ray observations
are capable of providing additional information~-- namely the fraction
and orientation of linearly polarized photons~-- and would help to
constrain different emission models\cite{Krawcz2011} of sources with
compact emission regions and high X-ray fluxes such as mass-accreting
black holes (BHs) and neutron stars. So far, only a few missions have
successfully measured polarization in the soft
(OSO-8\cite{Weisskopf1978}) and hard (Integral\cite{Dean2008}) X-ray
energy regime. The Crab nebula is the only source for which the
polarization of the X-ray emission has been established with a high
level of confidence \cite{Weisskopf1978, Dean2008}. The source exhibits
a polarization fraction of $20 \%$ at energies of $2.6 - 5.2 \,
\rm{keV}$ (direction angle of $30 \deg$ with respect to the X-ray
jet)\cite{Weisskopf1978} and $46 \% \pm 10 \%$ above $100 \, \rm{keV}$
(direction aligned with the X-ray jet observed in the nebula). Integral
observations of the X-ray binary Cygnus\,X-1 indicate a high fraction of
polarization in the hard X-ray/gamma-ray bands \cite{Laurent2011}. Model
predictions of polarization fraction for various source types lie
slightly below the sensitivity of the past OSO-8 mission, making future,
more sensitive polarimeter missions particularly interesting.

\paragraph{Future missions.} As polarimetry was not the main objective
of the Integral mission, the results are plagued by large systematic
uncertainties, and there are currently no other missions in orbit that
are capable of making sensitive X-ray polarimetric observations. This
will change by the launch of the satellite-borne {\it Gravity and
Extreme Magnetism SMEX} (GEMS) mission\cite{GEMS} scheduled for 2014.
GEMS will use two Wolter-type X-ray mirrors to focus $2 - 10 \,
\rm{keV}$ photons onto photo-effect polarimeters. For higher energies $E
> 10 \, \rm{keV}$ X-ray polarimeter designs usually make use of the
Compton effect: photons scatter preferentially perpendicular to the
orientation of the electric field vector~-- the azimuthal distribution
of scattered events will therefore show a sinusoidal modulation with
$180 \deg$ periodicity and a maximum at $\pm 90 \deg$ to the preferred
electric field direction of a polarized X-ray signal. The {\it Soft
Gamma-Ray Imager} on {\it ASTRO-H}\cite{Tajima2010} (launch scheduled
for 2013) will have capabilities of measuring polarization, but the
results may be plagued by similar systematic uncertainties as with the
Integral results. The hard X-ray polarimeter {\it X-Calibur} discussed
in this paper has the potential to cover the energy range above $10 \,
\rm{keV}$. Furthermore, X-Calibur combines a high detection efficiency
with a low level of background and has well-controlled systematic
errors. These features make it a particularly useful instrument for
astronomical X-ray polarimetry.

\begin{figure}
   \begin{center}
   \begin{minipage}[c]{0.49\textwidth}
     \includegraphics[width=1.0\textwidth]{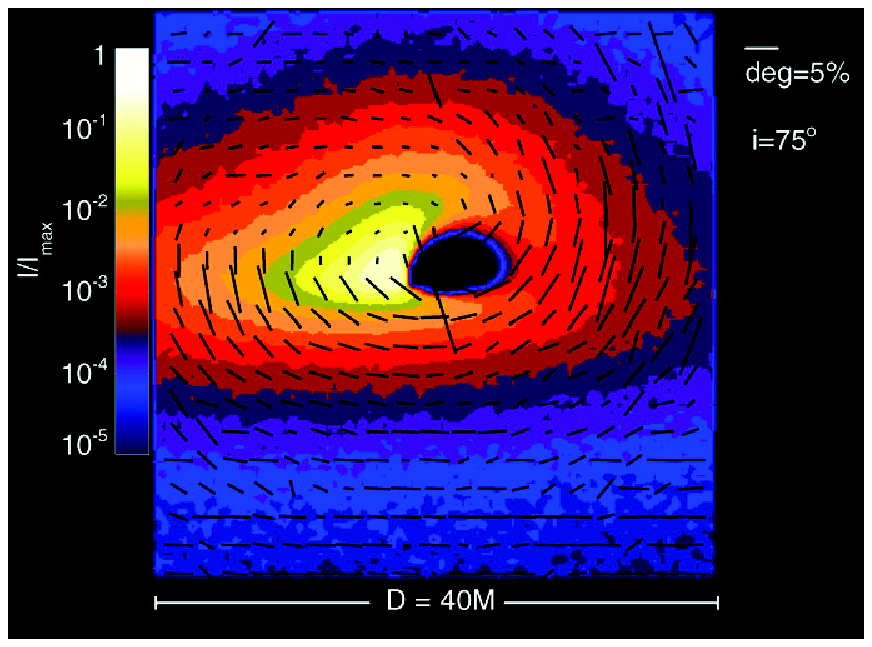}
   \end{minipage}
   \hfill
   \begin{minipage}[c]{0.45\textwidth}
     \includegraphics[width=1.0\textwidth]{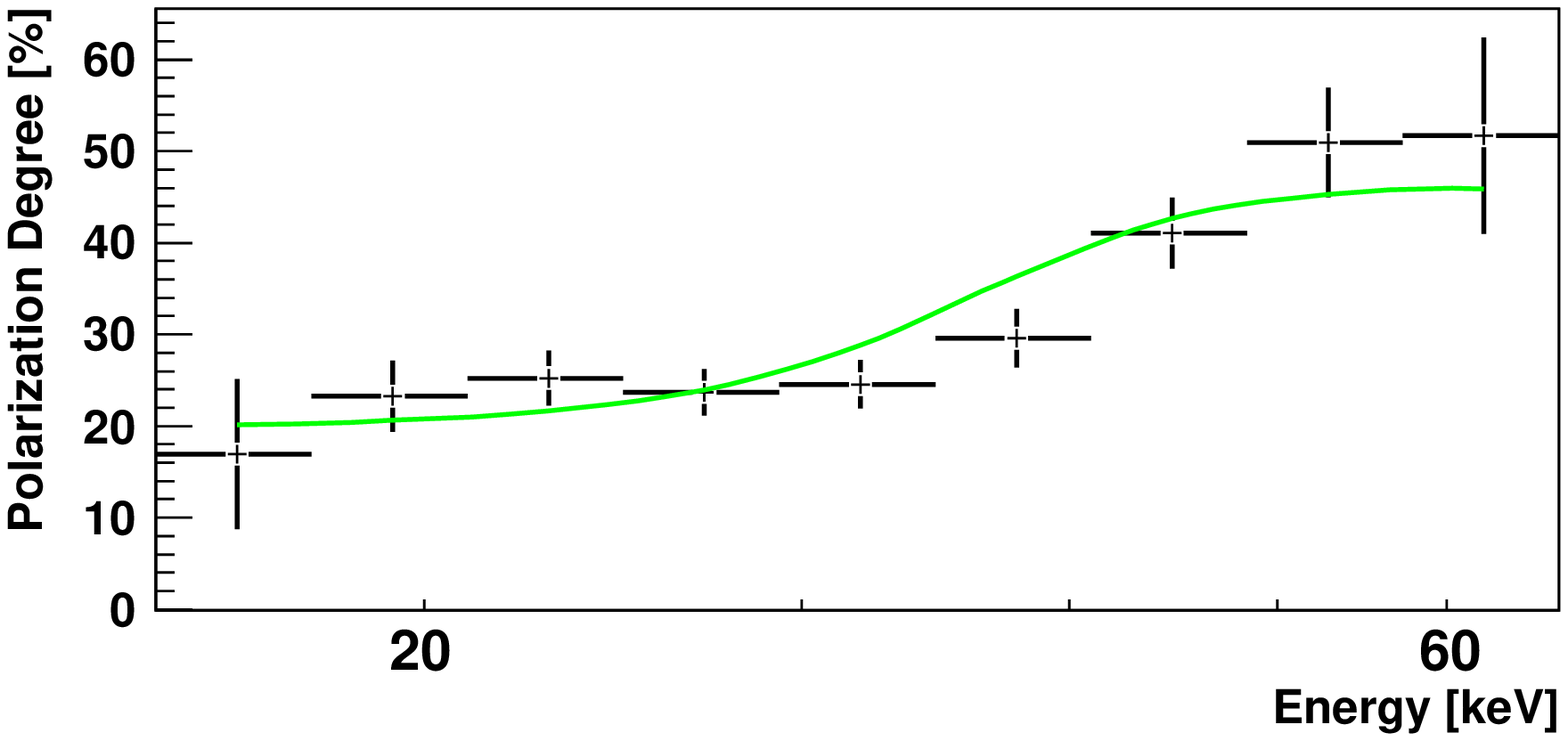}
     \includegraphics[width=1.0\textwidth]{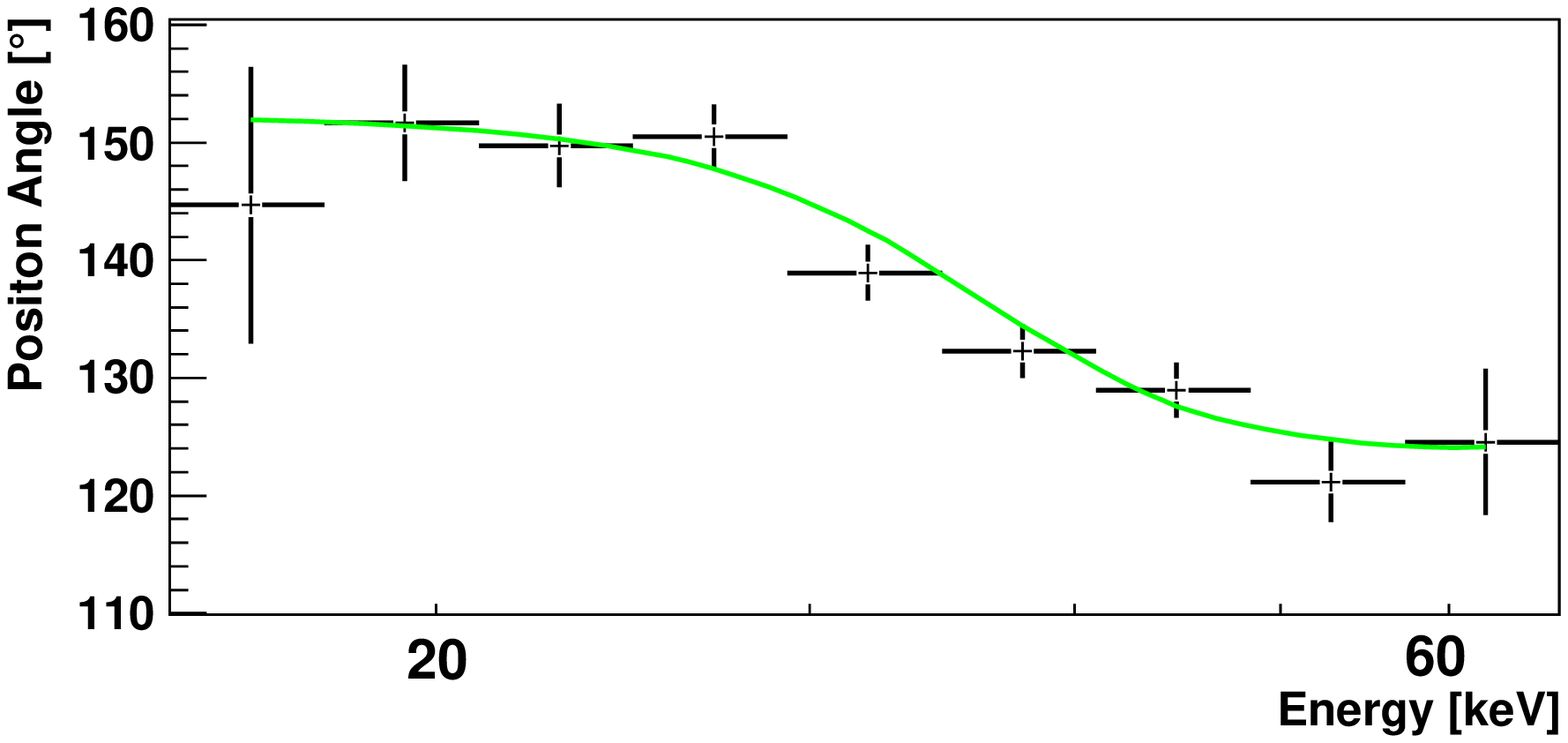}
   \end{minipage}
   \end{center}

\caption{\label{fig:BH_Polarization} {\bf Left:} Ray-traced image of
direct radiation from a thermal disk around a black hole including
returning radiation (observer located at an inclination angle of $75
\deg$, gas on the left side of the disk moving toward the observer
causing the characteristic increase in intensity due to relativistic
beaming). The observed intensity is color-coded on a logarithmic scale
and the energy-integrated polarization vectors are projected onto the
image plane with lengths proportional to the fraction of polarization.
The image has been adopted from Schnittman et al.
(2009)\cite{Schnittman2009}. {\bf Right:} Simulated X-Calibur
observations of the Crab ($5.6 \, \rm{h}$) as discussed in
Sec.~\ref{sec:simulations}. The data points show the reconstructed
fraction of polarization (top) and polarization angle (bottom) for the
assumed model (solid lines).}

\end{figure} 

\paragraph{Scientific potential.} Polarization measurements are of
general interest as tests of non-thermal emission processes in the
Universe. Synchrotron emission, for example, will result in linearly
polarized photons with their electric fields oriented perpendicular to
the magnetic field lines (projected) and the observed polarization map
in the X-rays can therefore be used to trace the magnetic field
structure of the source. An electron population with a spectral energy
distribution of $\rm{d}N/\rm{d}E \propto E^{-p}$ emitting in a uniform
magnetic field will lead to an observable fraction of
polarization\cite{Korchakov1962} of $f_{\rm{sync}} = (p+1)/(p+7/3)$ with
$(p + 3) = (\alpha + 1) / (\alpha + 5/3 )$, where $\alpha$ is the index
of the X-ray power spectrum. An observed polarization fraction close to
this limit can therefore be interpreted as an indication of a highly
ordered magnetic field since non-uniformities in the magnetic field will
reduce the fraction of polarization. The polarized synchrotron photons
can be inverse-Compton (IC) scattered by relativistic electrons~--
weakening the fraction of polarization (but not erasing it) and
imprinting a scattering angle dependence\cite{Poutanen1994} to the
observed fraction of polarization. Such IC signals usually (but not
always) appear in hard gamma-rays, where polarimetry is difficult, due
to multiple scattering in pair production detectors. Another important
mechanism for polarizing photons is Thomson scattering which creates a
polarization perpendicular to the scattering plane. Curvature radiation
is polarized, as well. The scientific potentials of spectro-polarimetric
hard X-ray observations are listed below:

\begin{itemize}

\item {\it Binary black hole systems.} Particle scattering within a
Newtonian accretion disk will lead to the emission of polarized X-rays. 
Relativistic aberration and beaming, gravitational lensing, and
gravitomagnetic frame-dragging will result in an energy-dependent
fraction of polarization\cite{Connors1977} since photons with higher
energies originate closer to the BH than the lower-energy photons.
Schnittman and Krolik \cite{Schnittman2009, Schnittman2010} calculate
the expected polarization signature including the effects of deflection
of photons emitted in the disk by the strong gravitational forces in the
regions surrounding the black hole and of the re-scattering of these
photons by the accretion disk. The resultant effect is a swing in the
polarization direction from being horizontal at low energies to being
vertical at high energies, i.e., parallel to the spin axis of the black
hole. Spectropolarimetric observations can therefore be used to
constrain the mass and spin of the BH\cite{Schnittman2009}, as well as
the inclination of the inner accretion disk and the shape of the corona
\cite{Schnittman2010} (Fig.~\ref{fig:BH_Polarization}, left).

\item {\it Pulsars and pulsar wind nebulae.} High-energy particles in
pulsar magnetospheres are expected to emit synchrotron and/or curvature
radiation which are difficult to distinguish from one another, solely
based on the observed photon energy spectrum. However, since the orbital
planes for accelerating charges that govern these two radiation
processes are orthogonal to each other, their polarized emission will
exhibit different behavior in position angle and polarization fraction
as functions of energy and the phase of the pulsar\cite{Dean2008}. In
magnetars, the highly-magnetized cousins of pulsars,
polarization-dependent resonant Compton up-scattering is a leading
candidate for generating the observed hard X-ray tails (e.g. Baring \&
Harding 2007 \cite{BaringHarding2007}). In both these classes of neutron
stars, phase-dependent spectropolarimetry can probe the emission
mechanism, and provide insights into the magnetospheric locale of the
emission region. Furthermore, spectropolarimetric observations can be
used to constrain the magnetic field and particle populations in pulsar
wind nebulae such as the Crab, the leading driver for this field of
X-ray polarimetry. These objects potentially show a higher polarization
fraction at hard X-rays as compared to soft X-rays, reflecting the
contrast between jet and more diffuse nebular contributions.

\item {\it Relativistic jets in active galactic nuclei.} Relativistic
electrons in jets of active galactic nuclei (AGN) emit polarized
synchrotron radiation at radio/optical wavelengths. The same electron
population is believed to produce hard X-rays by inverse-Compton
scattering of a photon field. Simultaneous measurements of the
polarization angle and the fraction of polarization in the radio to hard
X-ray band could help to address the following questions: (i) If the
electrons mainly up-scatter the co-spatial synchrotron photon field
(synchrotron self Compton), the polarization of the hard X-rays is
expected to track the polarization at radio/optical
wavelengths\cite{Poutanen1994}. The fraction of polarization could be a
substantial part of the synchrotron fraction of polarization and the
polarization directions should be identical. (ii) If the electrons
dominantly up-scatter an external photon field (external Compton, e.g.
photons of the cosmic microwave background) the hard X-rays will have a
relatively small ($<$10\%) fraction of polarization\cite{McNamara2009}. 
Polarization also allows one to test the structure of the magnetic field
of the jet: Particles accelerated in a helical field which are moving
through a standing shock can cause an X-ray synchrotron flare with a
continuous (in time) swing in polarization direction. Such an event was
recently observed from BL\,Lacertae at optical wavelengths
\cite{Marscher2008} and could potentially be observable at X-ray
energies, as well.

\item {\it Gamma-ray bursts}. Gamma-ray bursts are believed to be
connected to hyper-nova explosions and the formation/launch of
relativistic jets\cite{Woosley1993}.  As in the case of the jets in AGN,
the structure and particle distribution responsible for gamma-ray bursts
will be revealed by X-ray polarization measurements. The X-ray emission
of a gamma-ray burst usually lasts for only a few minutes, so that rapid
follow-up observations in the X-ray band would be the main challenge.
 
\item {\it Lorentz invariance.} Hard X-ray polarimetric observations can
be used to test/constrain some theories violating Lorentz
invariance\cite{Fan2007} with unprecedented accuracies by probing the
helicity dependence of the speed of light.

\end{itemize}

For more details on the scientific prospects see for
example\cite{Krawcz2011, Lei1997} and references therein. Addressing
these science goals requires spectro-polarimetric observations over the
broadest possible energy range.

\paragraph{Definitions.} If we assume a $100 \%$ linearily polarized
photon beam, then the minimum/maximum number of counts
$C_{\rm{min}}$/$C_{\rm{max}}$ of the azimuthal Compton-scattering
distribution (histogram) can be used to define the modulation factor:
$\mu = (C_{\rm{max}} - C_{\rm{min}}) / (C_{\rm{max}} + C_{\rm{min}})$.
It represents the modulation amplitude of a $100 \%$ polarized beam and
depends on the polarimeter design and the physics of Compton-scattering.
The performance of a polarimeter can be characterized by the minimum
detectable polarization (MDP) as the minimum fraction of polarization
that can be detected at the $99 \%$ confidence level. Assuming a
polarimeter that detects all Compton-scattered photons and has an ideal
angular resolution~-- in this case $\mu$ becomes the modulation
amplitude averaged over all solid angles and the Klein-Nishina cross
section~-- one can estimate the MDP by integrating the scattering
probability distribution \cite{KrawczAnalysisPOL}:

\begin{equation}
\label{eq:MDP}
\rm{MDP} \simeq \frac{4}{\mu R_{\rm{src}}} \sqrt{\frac{R_{\rm{src}} + 
R_{\rm{bg}}}{T}}
\end{equation}

$T$ is the observation time, $R_{\rm{src}}$ and $R_{\rm{bg}}$ are the
source and background count rates, respectively.

\section{DESIGN OF X-CALIBUR}
\label{sec:design}

\begin{figure}[t]
   \begin{center}
   \begin{tabular}{c}
   \includegraphics[width=0.8\textwidth]{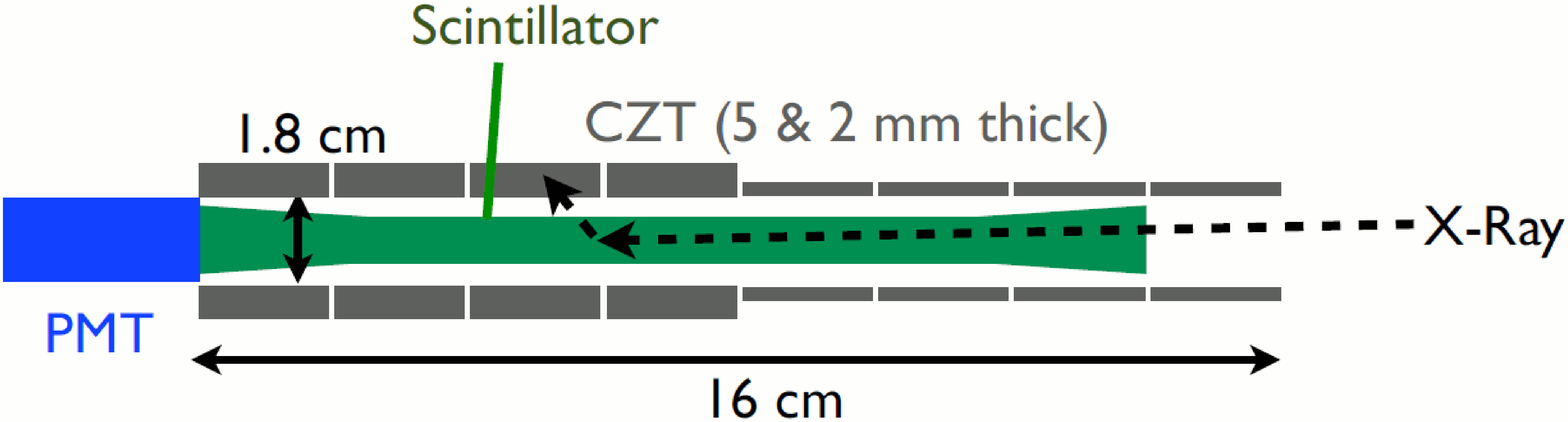} \\
   \includegraphics[width=0.42\textwidth]{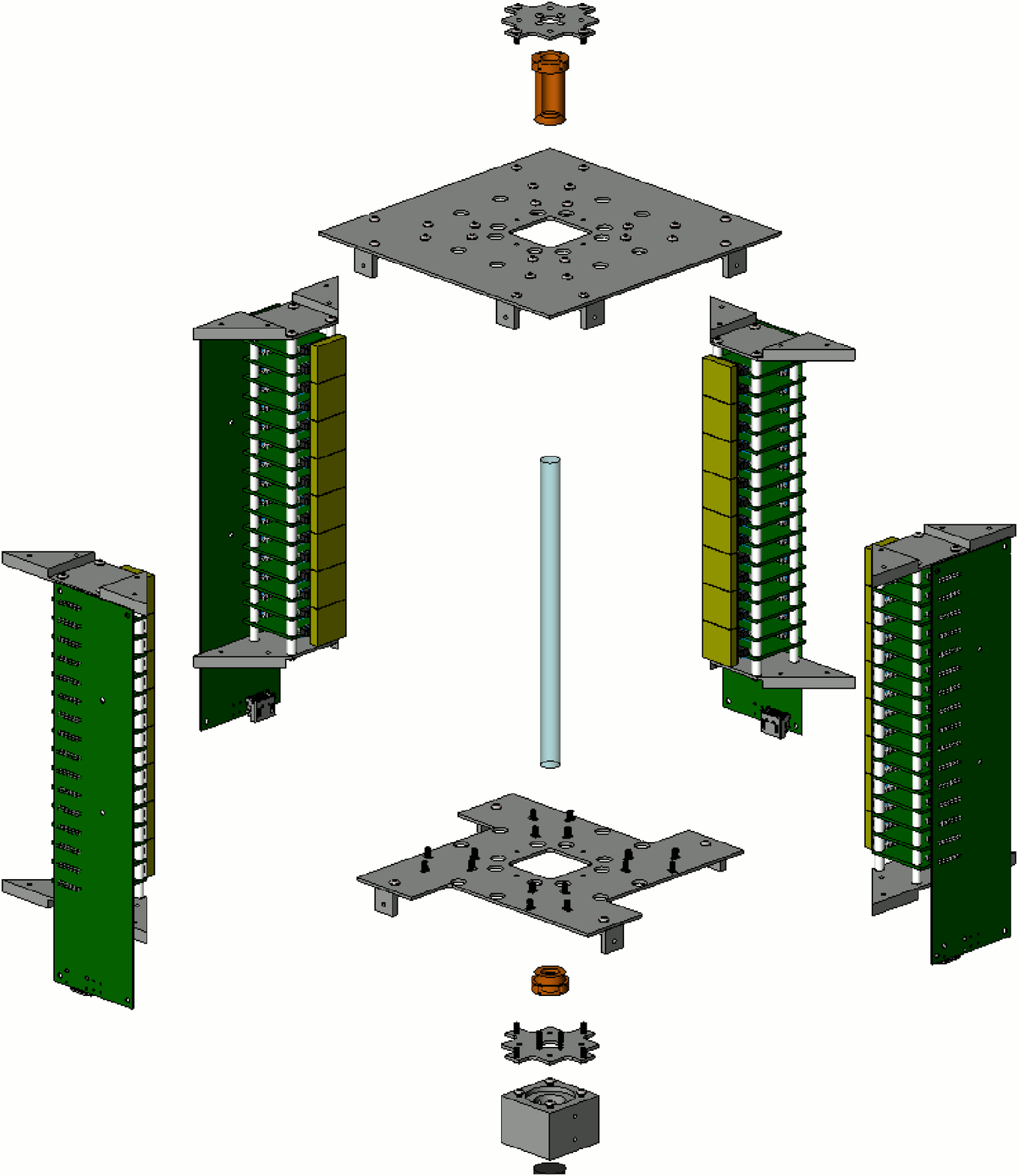}
   \hfill
   \includegraphics[width=0.26\textwidth]{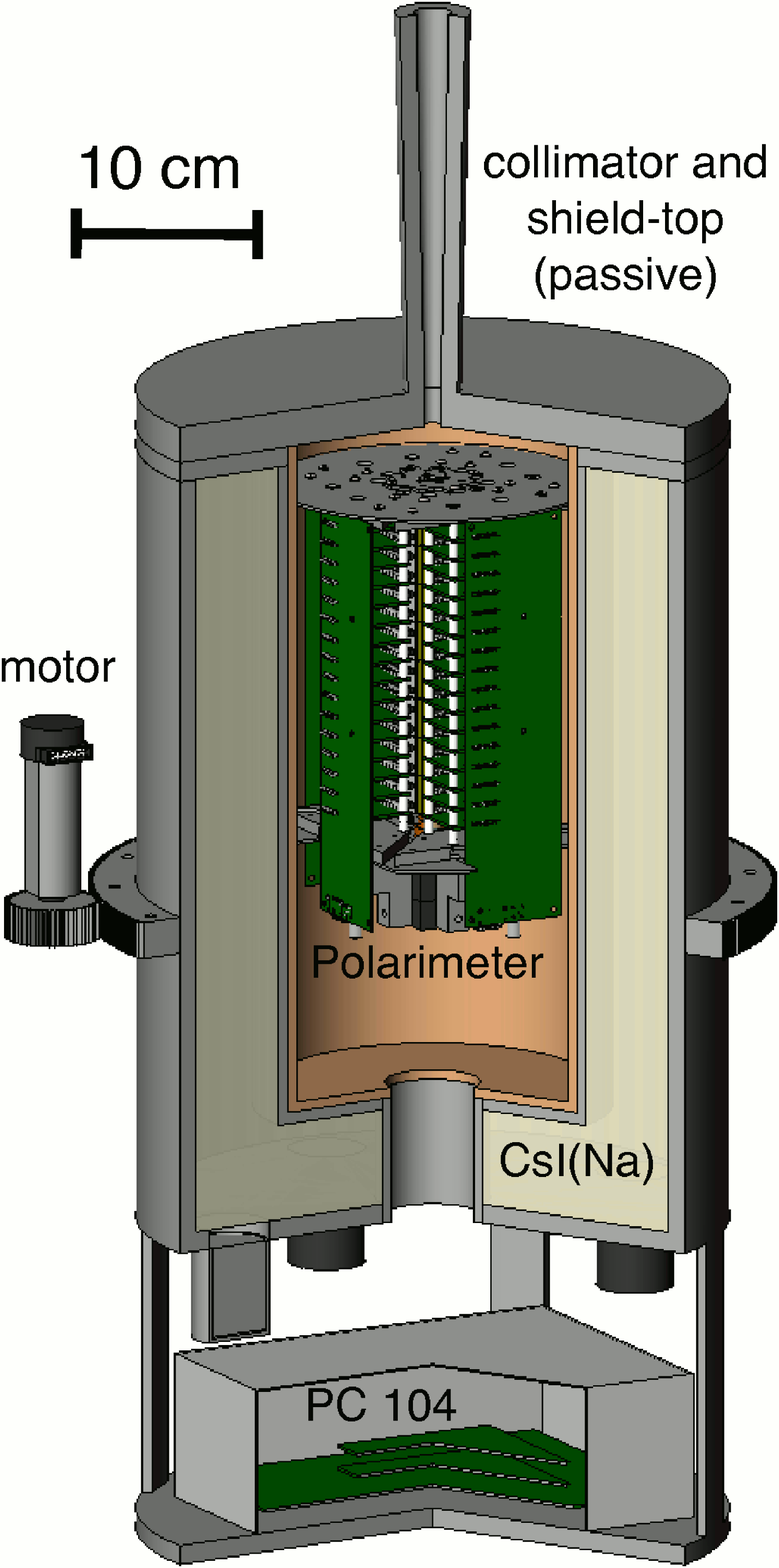}
   \hfill
   \includegraphics[width=0.28\textwidth]{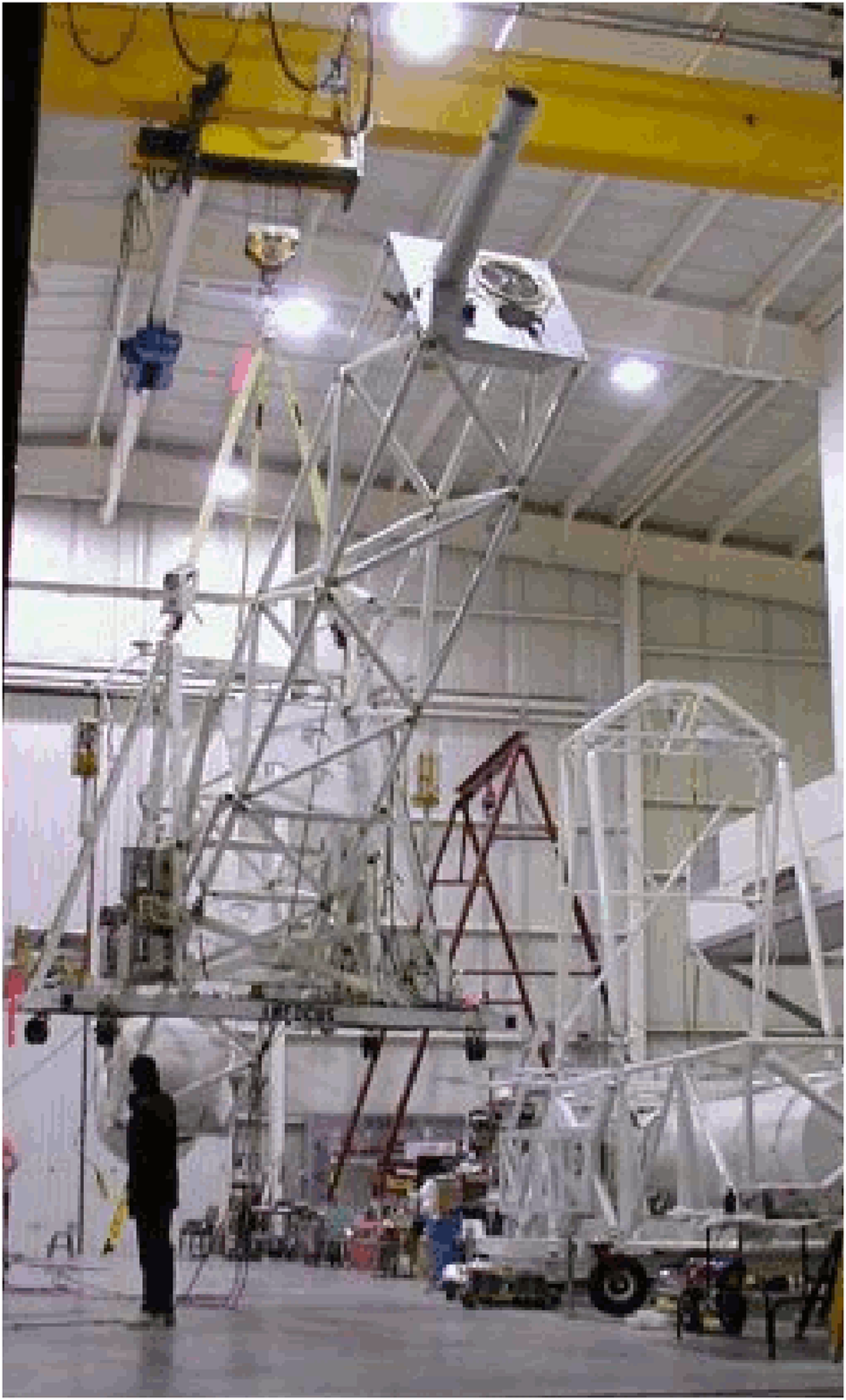}
   \end{tabular}
   \end{center}

\caption{\label{fig:Design} {\bf Top:} Conceptual design of X-Calibur:
Incoming X-rays are Compton-scattered in the scintillator rod (read by a
PMT) and are subsequently photo-absorbed in a CZT detector. {\bf Lower
left:} View of the X-Calibur polarimeter (``exploded view''): 4 sides of
CZT detector columns surround the central scintillator rod (optical
axis). {\bf Lower middle:} X-Calibur with shielding and azimuthal
rotation bearing. {\bf Lower right:} InFOC$\mu$S balloon gondola.
X-Calibur will be situated in the focal plane~-- roughly $8 \, \rm{m}$
away from the Wolter X-ray mirror ($40 \, \rm{cm}$ diameter).}

\end{figure} 

The conceptual design of the X-Calibur polarimeter is illustrated in the
top panel of Figure~\ref{fig:Design}. A low-Z scintillator is used as
Compton-scatterer for which the cross-section of the photo-effect can be
neglected as compared to the cross-section of Compton-scattering for
energies $> 15 \, \rm{keV}$ (defining the low-energy threshold of
X-Calibur). The length of the scintillator rod is chosen such that even
$80 \, \rm{keV}$ photons Compton-scatter with a probability of $90 \%$.
For sufficiently energetic photons, the Compton interaction produces a
measurable scintillator signal which is read by a PMT. The scattered
X-rays are photo-absorbed in surrounding rings of high-Z Cadmium Zinc
Telluride (CZT) detectors. This combination of scatterer/absorber leads
to a high fraction of unambiguously detected Compton events. Linearly
polarized X-rays will preferably Compton-scatter perpendicular to their
E field vector~-- resulting in a modulation of the azimuthal event
distribution (see Sec.~\ref{sec:intro}). 

\begin{figure}
   \begin{center}
   \begin{tabular}{c}
   \includegraphics[width=0.31\textwidth]{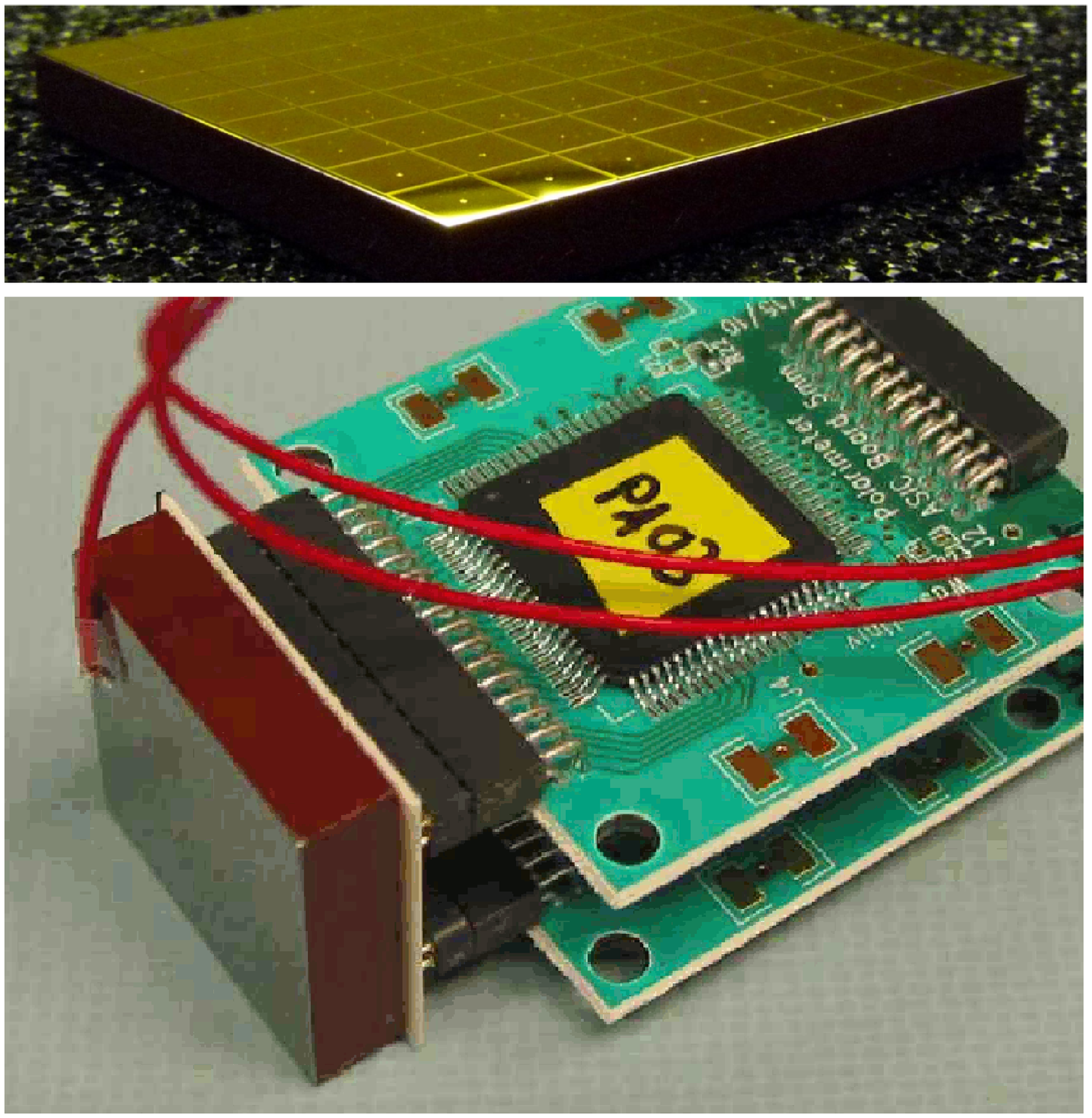}
   \hfill
   \includegraphics[width=0.21\textwidth]{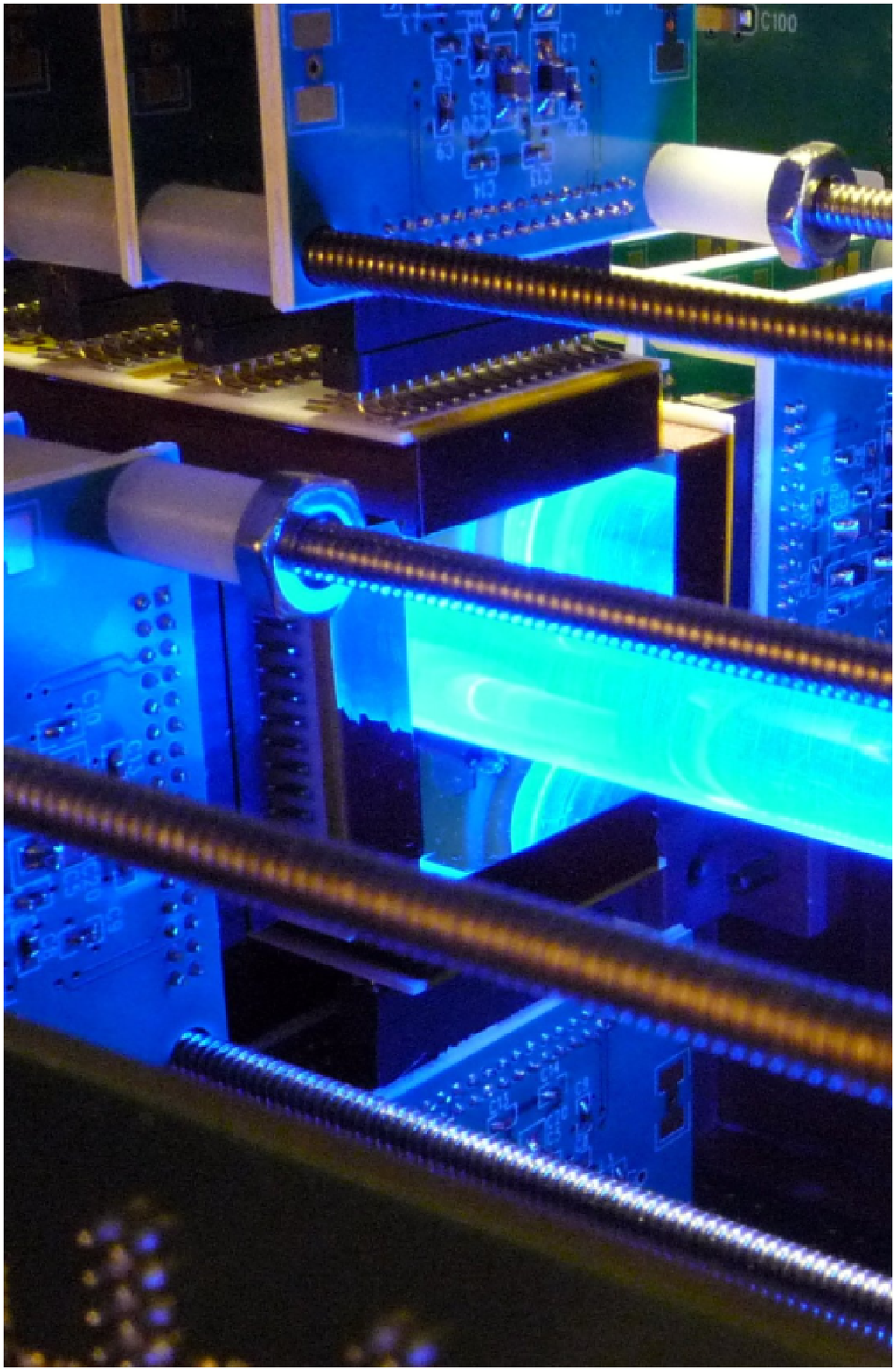}
   \hfill
   \includegraphics[width=0.46\textwidth]{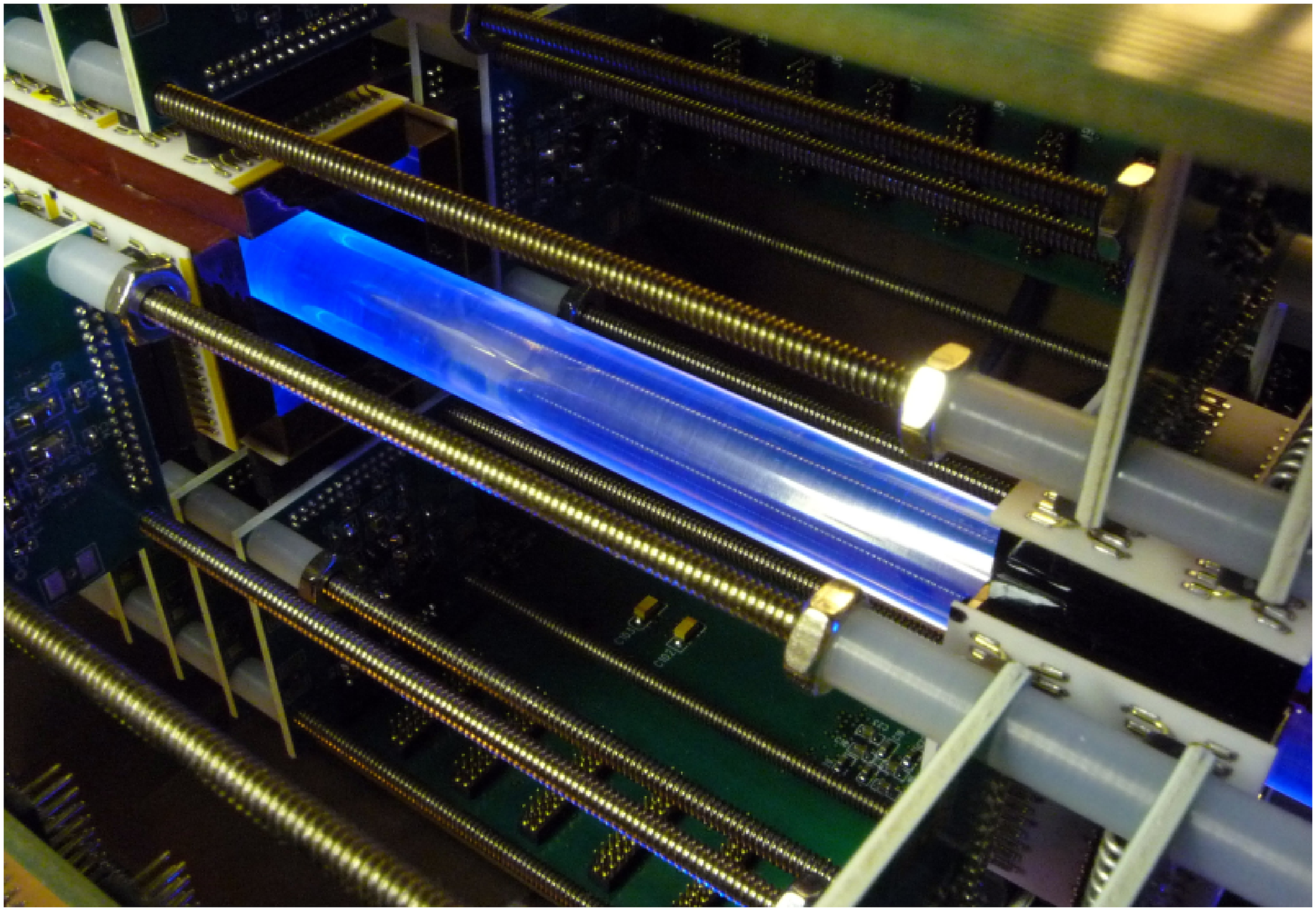}
   \end{tabular}
   \end{center}

\caption{\label{fig:Fotos} {\bf Left:} Top: $2 \times 2 \times 0.2 \,
\rm{cm}^{3}$ CZT detector with 64 pixels (anode). Bottom: $2 \times 2
\times 0.5 \, \rm{cm}^{3}$ CZT detector bonded to a ceramic chip carrier
which is plugged into 2 ASIC readout boards. The high-voltage cable is
glued to the detector cathode (red wire). {\bf Middle:} The scintillator
(blueish glow) being surrounded by two detector rings~-- each consisting
of four 64 pixel CZT detectors. {\bf Right:} X-Calibur equipped with
three detector rings (2 on the left and one on the right).}

\end{figure}

The CZT detectors were ordered from different companies (Endicott
Interconnect, Quikpak/Redlen, Creative Electron). Each detector ($2
\times 2 \, \rm{cm}^{2}$) is contacted with a 64-pixel anode grid ($2.5
\, \rm{mm}$ pixel pitch) and a monolithic cathode facing the
scintillator rod. Two detector thicknesses ($0.2 \, \rm{cm}$ and $0.5 \,
\rm{cm}$) are being tested in the current setup. Each CZT detector is
permanently bonded (anode side) to a ceramic chip carrier which can be
plugged into the electronic readout board. Figure~\ref{fig:Fotos} (left)
shows a single CZT detector unit with a $8 \times 8$ pixel matrix on the
anode side as well as the readout electronics. Each CZT detector is read
out by two digitizer boards (32 channel ASIC developed by G.~De~Geronimo
(BNL) and E.~Wulf (NRL) \cite{Wulf2007} and a 12-bit ADC). The readout
noise is as low as $2.5 \, \rm{keV}$ FWHM. 16 digitizer boards (8 CZT
detectors) are read out by one harvester board transmitting the data to
a PC-104 computer with a rate of 6.25~Mbits/s. X-Calibur will comprise
2048 data channels. Four detector units form a 'ring' surrounding the
scintillator slab. The scintillator EJ-200 (H:C $=5.17:4.69$, $\left< Z
\right> = 3.4$, $\rho = 1.023 \, \rm{g} / \rm{cm}^{3}$, decay time $2.1
\, \rm{nsec}$) is read by a Hamamatsu R7600U-200 PMT with a high quantum
efficiency super-bi-alkali photo cathode. The PMT trigger information
allows to effectively select scintillator/CZT events from the data which
represent likely Compton-scattering candidates. The polarimeter and the
front-end readout electronics will be located inside an active CsI(Na)
anti-coincidence shield with $5 \, \rm{cm}$ thickness and a passive lead
shield/collimator at the top (Fig.~\ref{fig:Design}) to suppress charged
and neutral particle backgrounds.

We plan to use the X-Calibur polarimeter in the focal plane of the
InFOC$\mu$S\footnote{{\tt http://infocus.gsfc.nasa.gov/}} experiment
\cite{InFocus_FirstFlight} (Fig.~\ref{fig:Design}, right).  The total
mass of the gondola and the X-ray telescope will be $1,400 \, \rm{kg}$.
A Wolter grazing incidence mirror\footnote{Grazing incidence mirrors
change the polarization of the X-ray photons by less than $1 \%$
\cite{Katsuta2009}} focuses the X-rays on the polarimeter. The X-Calibur
scintillator rod will be aligned parallel to the optical axis of the
InFOC$\mu$S X-ray telescope. The focal length is $\sim 8 \, \rm{m}$ and
the field of view (FWHM) is $10 \, \rm{arcmin}$\footnote{Note, that the
X-Calibur polarimeter does not provide imaging capabilities.}. The
design of InFOC$\mu$S allows for very stable pointing of the telescopes
to $<$1 arcmin as the focus of the X-ray telescope moves across the sky. 
In order to reduce the systematic uncertainties of the polarization
measurements (including biases generated by the active shield, a
possible pitch of the polarimeter with respect to the X-ray telescope,
etc.), the polarimeter and the active shield will be rotated around the
optical axis ($\sim 10 \, \rm{rpm}$) using a ring bearing (see
Fig.~\ref{fig:Design}, middle). Counter-rotating masses will be used to
cancel the net-angular momentum of the polarimeter assembly during the
flight. Power will be provided to the polarimeter by sliding contacts
and communication will be done via a wireless network. The data will be
stored to solid state drives and will be down-linked to the ground. The
advantages of the X-Calibur/InFOC$\mu$S design are (i) a high detection
efficiency by using more than $80 \%$ of photons impinging on the
polarimeter, (ii) low background due to the usage of a focusing optics
instead of large detector volumes, and (iii) minimization and control of
systematic effects and achievement of a corresponding quantitative
estimate thereof.

\section{SIMULATIONS}
\label{sec:simulations}

Simulations of the X-Calibur polarimeter were performed using the {\it
Geant4} package\footnote{{\tt http://geant4.cern.ch/}} with the
Livermore low-energy electromagnetic model list. All relevant/important
effects are included in the simulation starting from the detailed
experimental design (see Sec.~\ref{sec:design} and
Fig.~\ref{fig:Design}), the shielding, the X-ray telescope, as well as
several backgrounds. A balloon flight in the focal plane of the
InFOC$\mu$S mirror assembly was assumed. The effective detection areas
of the X-ray mirror are $95/60/40 \, \rm{cm}^{2}$ at $20/30/40 \,
\rm{keV}$. We accounted for atmospheric absorption at a floating
altitude of $130,000$ feet using the NIST XCOM attenuation
coefficients\footnote{{\tt http://www.nist.gov/pml/data/xcom/index.cfm}}
and an atmospheric depth of $2.9 \, \rm{g/cm}^{2}$ (observations
performed at zenith)~-- the atmospheric transmissivity rapidly increases
from $0$ to $0.6$ in the $20 - 60 \, \rm{keV}$ range. We simulated the
most important backgrounds such as the cosmic X-ray
background\cite{Ajello2008}, albedo photons and cosmic ray protons and
electrons\cite{Mizuno2004}. The neutron background was not modeled since
a detailed study of Parsons et al. (2004) showed that the background
contribution in CZT crystals can be neglected\cite{Parsons2004}. A
Crab-like source was simulated for a $5.6 \, \rm{hr}$ balloon flight. We
assumed a power-law energy spectrum, and a continuous change of the
fraction of polarization and the polarization angle between the values
measured at $5.2 \, \rm{keV}$ with OSO-8\cite{Weisskopf1978} and at
$E>100 \, \rm{keV}$ with Integral\cite{Dean2008} by modeling a
transition following a Fermi distribution
(Fig.~\ref{fig:BH_Polarization}, right). The simulation data were
analyzed in the same way as the experimental data.

\begin{figure}
   \begin{center}
   \begin{tabular}{c}
   \includegraphics[width=0.49\textwidth]{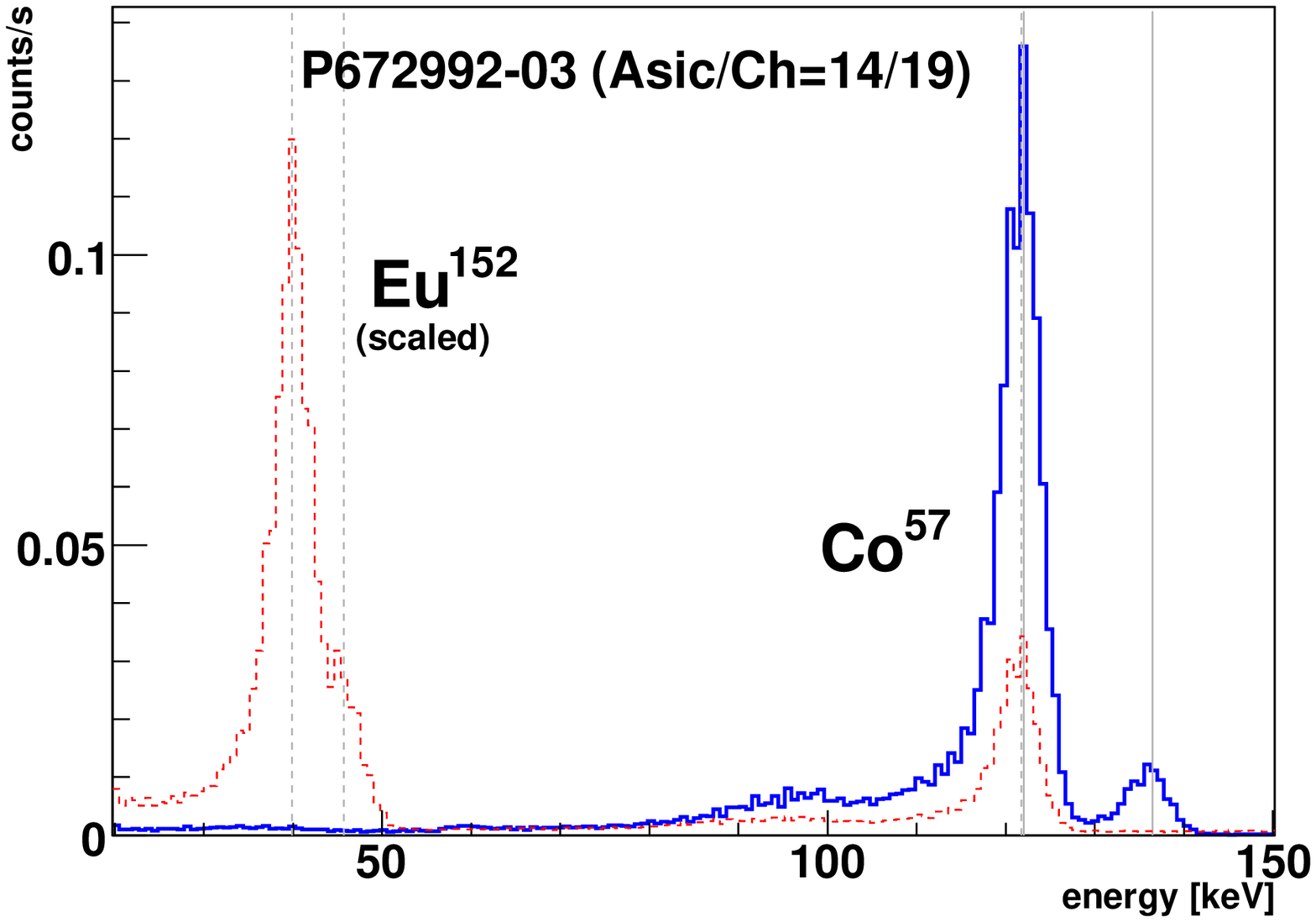}
   \includegraphics[width=0.49\textwidth]{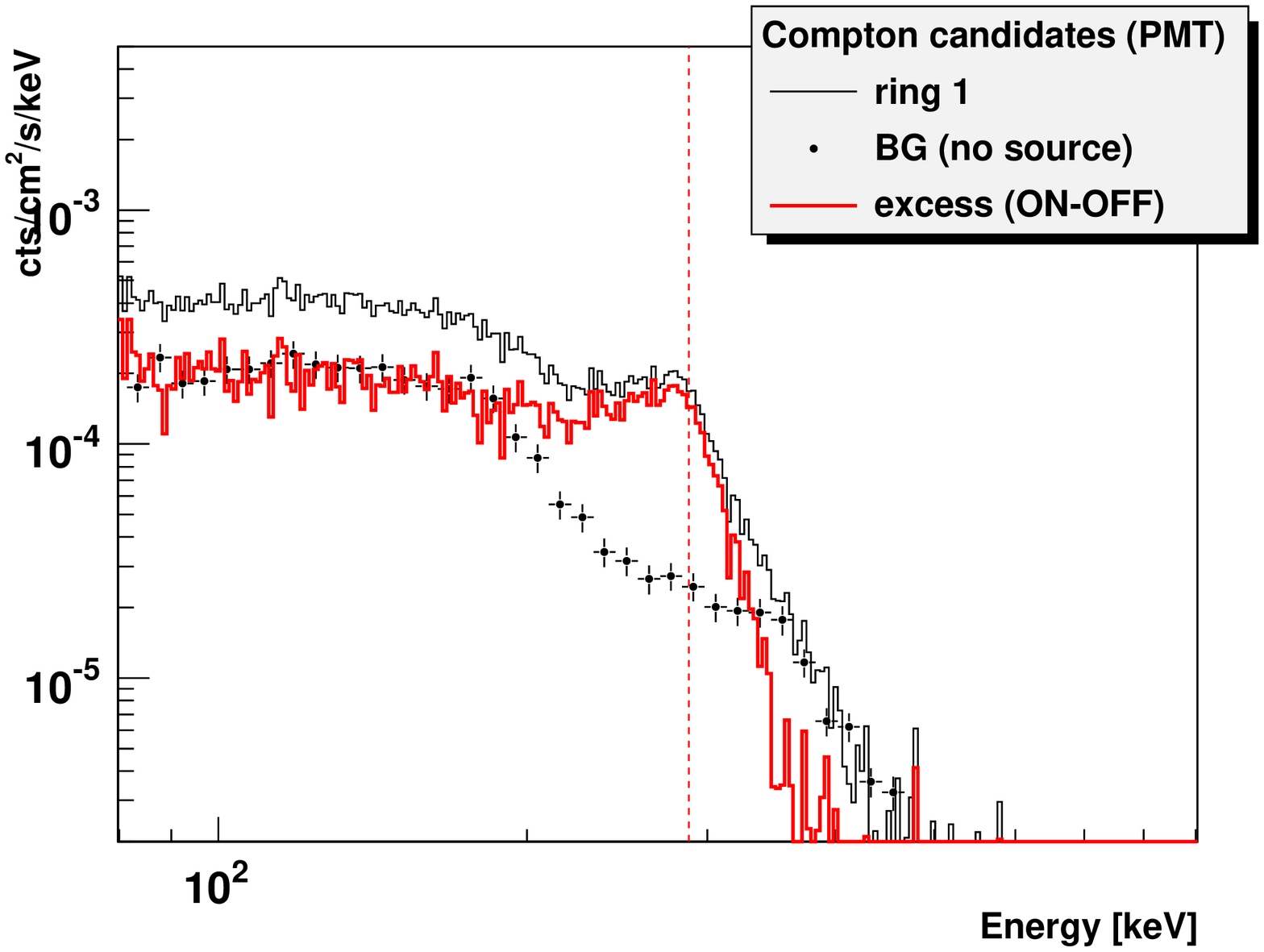}
   \end{tabular}
   \end{center}

\caption{\label{fig:Spectra} {\bf Left:} Energy spectra of a Eu$^{152}$
source (dashed, scaled) and a Co$^{57}$ source (solid) measured with one
pixel of a calibrated CZT detector. The vertical lines indicate the
nominal X-ray line energies (from left to right: $39.9 \, \rm{keV}$,
$45.7 \, \rm{keV}$, $122.1 \, \rm{keV}$ and $136.5 \, \rm{keV}$). {\bf
Right:} Energy spectrum of $288 \, \rm{keV}$ X-rays (partially
polarized) after being scattered in the scintillator (different
scattering angles and corresponding energy transfer). The background
measurement is done without a source (cosmic rays). The vertical line
($288 \, \rm{keV}$) indicates the energy of the incoming X-ray beam.}

\end{figure} 

For a Crab-like source the simulations predict an event rate of $1.1
(3.2) \, \rm{Hz}$ with (without) requiring a triggered scintillator
coincidence detected by the PMT. The X-Calibur modulation factor is $\mu
= 0.52$ for a $100 \%$ polarized beam. The MDP (see \ref{eq:MDP}) in the
$10-80 \, \rm{keV}$ energy range will be $4 \%$ assuming $5.6 \,
\rm{hr}$ of on-source observations of a Crab-like source combined with a
1.4~hr background observation of an adjacent empty field. The expected
background rates are $0.04 (1.1) \, \rm{Hz}$ with (without) requiring a
scintillator coincidence. Different shield configurations and shield
thicknesses were simulated. The configuration shown in
Fig.~\ref{fig:Design} (middle) represents an optimized compromise
balancing the background rejection power and the mass/complexity of the
shield.

The results of the simulated X-Calibur observations of the Crab Nebula
are compared to the assumed model curves in
Figure~\ref{fig:BH_Polarization}, right; the errors were computed in a
similar way as described by Weisskopf et al. (2010)
\cite{Weisskopf2010}. Simulations performed at different zenith angles
$\theta$ show that the sources rate scales with $(\cos \theta)^{1.3}$
which is taken into account for simulating astrophysical observations.
More detailed simulations are discussed in Guo et al.
(2010)\cite{Guo2010}. Krawczynski et al. \cite{Krawcz2011} present
MonteCarlo based comparisons between the performance of X-Calibur with
the performance of competing hard X-ray polarimeter designs in the $20 -
100 \, \rm{keV}$ energy range. X-Calibur outperforms the alternative
designs by a wide margin.

\section{FIRST MEASUREMENTS}
\label{sec:FirstMeasurements}

Using funding from Washington University's McDonnell Center for the
Space Sciences, a flight-ready version of the X-Calibur polarimeter was
assembled and tested in the laboratory. First measurements were
performed with 3 detector rings installed ($0.5 \, \rm{cm}$ thickness),
comprising a total of $3 \times 4 = 12$ detectors ($768$ data channels).
Before installation, IV-curves were taken for all pixels of the
detectors, followed by a calibration run using a Eu$^{152}$ source
(X-ray lines at $39.9$, $45.7$, $121.8$ and $344.3 \, \rm{keV}$). An
example of the calibration spectrum for one particular detector pixel is
shown in Fig.~\ref{fig:Spectra}, left. The energy resolution for this
particular detector is $4.1 \, \rm{keV}$ at $40 \, \rm{keV}$ and $5.0 \,
\rm{keV}$ at $121.8 \, \rm{keV}$. After calibration, a collimated
Eu$^{152}$ source was placed in front of the entrance window of the
polarimeter in order to determine the azimuthal detector acceptance for
an unpolarized beam. Only CZT events with a simultaneous ($30 \mu
\rm{s}$) scintillator trigger are used for this and the following
analysis; this is a very effective cut for selecting events of photons
that Compton-scattered in the scintillator rod and were subsequently
absorbed in one of the CZT detectors. Event rates were normalized by the
azimuthal angle $\Delta \Phi$ covered by the corresponding pixel.
Another data run was taken without any source to determine the
background induced by cosmic rays secondaries hitting the detector
assembly.

A polarized beam was generated by scattering a strong Cs$^{137}$ source
(line at $662 \, \rm{keV}$) off a lead brick. A lead collimator allowed
only X-rays with a scattering angle of $\sim 90 \deg$ to enter the
polarimeter. The X-ray beam of the scattered photons has a mean energy
of $288 \, \rm{keV}$ and was polarized to $\sim 55 \%$ (modulation
factor of $\mu = 0.4$). Therefore, the expected relative amplitude in
the normalized $\Phi$ distribution is $0.55 \times 0.4 = 0.22$.

Figure~\ref{fig:Spectra} (right) shows the raw spectrum of the first CZT
polarimeter ring measured from (i) the polarized beam, (ii) a background
spectrum measured without a source and (iii) the excess spectrum
corresponding to the energy spectrum of the scattered/polarized beam. As
expected, the excess spectrum drops off for energies higher than $288 \,
\rm{keV}$ (vertical line)~-- the energy of the $90 \deg$-scattered
Cs$^{137}$ photons entering the polarimeter. The continuum below this
energy is the result of $288 \, \rm{keV}$ photons being
Compton-scattered at different depths in the scintillator rod and
therefore being reflected to the first CZT ring under different
scattering angles and corresponding different Compton energy losses. The
little bump in the spectrum around $288 \, \rm{keV}$ may originate from
direct CZT hits without a Compton-scattering in the scintillator.

\begin{figure}
   \begin{center}
   \begin{tabular}{c}
   \includegraphics[width=\textwidth]{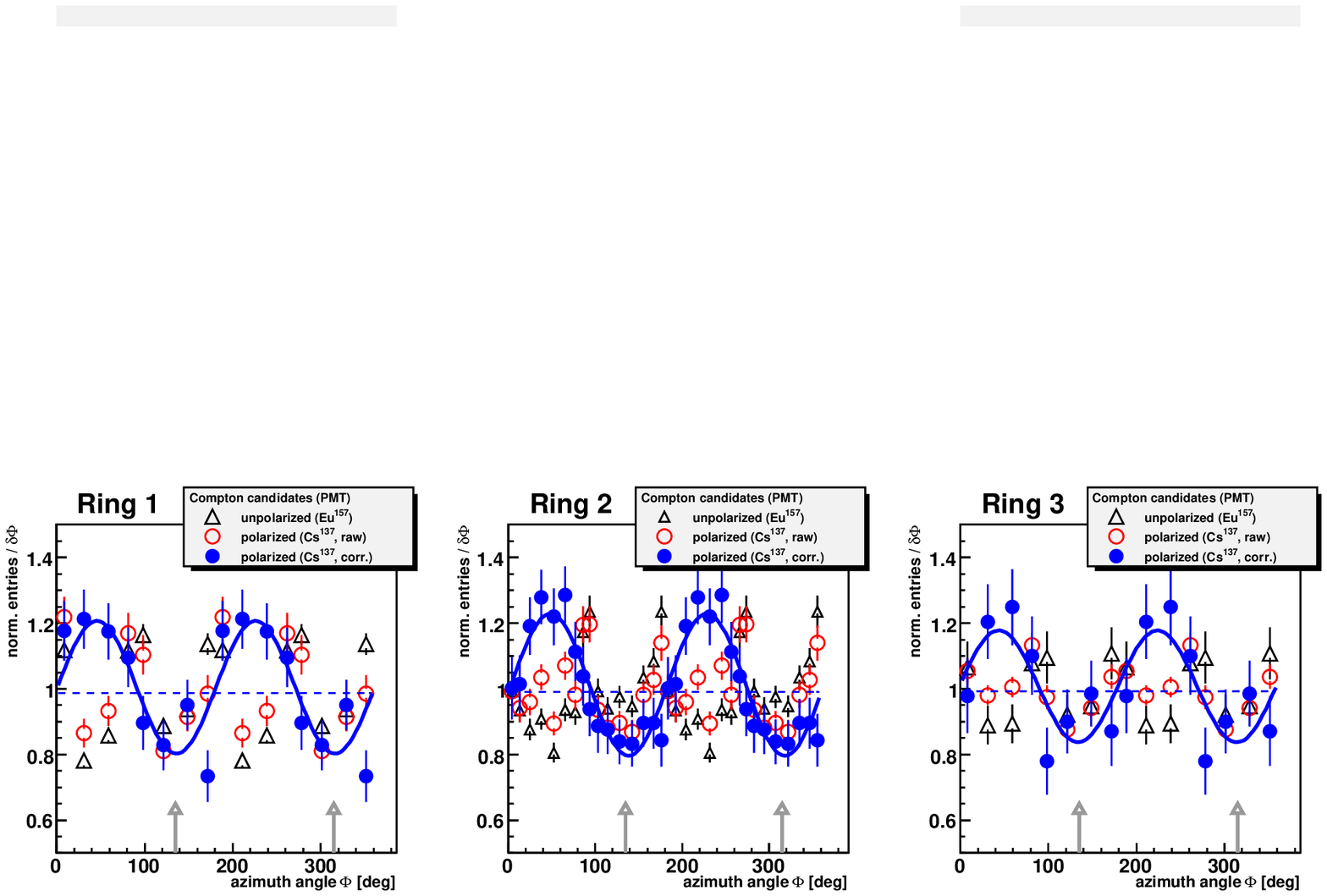}
   \end{tabular}
   \end{center}

\caption{\label{fig:PolarizationResults} Azimuthal event distribution
for the 3 installed CZT detector rings. Shown are raw events of the
polarized beam (red), an unpolarized beam (black), and the acceptance
corrected polarized events (blue). A sine function ($180 \deg$
modulation) was fit to the corrected data of the polarized beam. The
vertical arrows indicate the plane of the electric field vector. The $90
\deg$ modulation of the unpolarized beam is a result of the 4-fold
geometry of the CZT detector assembly. Ring~1 got less events than
ring~2 since it only sees backscatter-events (scintillator starts at
ring~2, see Fig.~\ref{fig:Design}), ring~3 got less events since it was
not included in the data acquisition at the start of data taking.}

\end{figure}

Figure~\ref{fig:PolarizationResults} shows the azimuthal scattering
distribution of the polarized and unpolarized beam for the three
installed CZT detector rings.  Only events with a simultaneous
scintillator trigger and with a deposited CZT energy between $100-330 \,
\rm{keV}$ are used (see spectrum in Fig.~\ref{fig:Spectra}, right). The
data of the polarized beam are corrected for the acceptance of the
polarimeter (derived from the unpolarized X-ray beam).  As expected for
a polarized beam, a $180 \deg$ modulation can be seen with a maximum of
azimuthal scattering angle perpendicular ($\Phi + 90 \deg$) to the plane
of the $E$ field vector of the polarized beam (indicated by the gray
arrows). A sine function was fit to the $\Phi$-distribution of the
polarized beam resulting in a relative amplitude of $0.22$. The data are
in excellent agreement with expectations.

\section{SUMMARY AND CONCLUSION}
\label{sec:summary}

We designed a hard X-ray polarimeter X-Calibur and studied its projected
performance and sensitivity for a 1-day balloon flight with the
InFOC$\mu$S X-ray telescope. X-Calibur combines a detection efficiency
of close to $100\%$ with a high modulation factor of $\mu \approx 0.5$,
as well as a good control over systematic effects. Compared to competing
designs of hard X-ray polarimeters X-Calibur does not only have superior
sensitivity but also a better energy resolution. X-Calibur was
successfully tested/calibrated in the laboratory with a polarized beam
of $288 \, \rm{keV}$ photons. Further laboratory measurements are
planned to study the effect of different orientations between the
polarization plane and the detector, the performance of different CZT
detector thicknesses ($0.2 \, \rm{cm}$ vs $0.5 \, \rm{cm}$), the active
shielding, and the rotation mechanism.

We applied for a 1-day X-Calibur/InFOC$\mu$S balloon flight in spring
2013. Our tentative observation program includes galactic sources (Crab,
Her\,X-1, Cyg\,X-1, GRS\,1915, EXO\,0331) and one extragalactic source
(Mrk\,421) for which sensitive polarization measurements would be
carried through. We envision follow-up longer duration balloon flights
(from the northern and southern hemisphere), possibly using a mirror
with increased area. An increased mirror area would lead to an increased
signal rate while leaving the background rate almost unchanged~--
resulting in an improved signal-to-noise ratio. In the ideal case these
flights would be performed while the GEMS mission is in orbit to achieve
simultaneous coverage in the $0.5 - 80 \, \rm{keV}$ regime. Successful
balloon flights would motivate a satellite-borne hard X-ray polarimetry
mission.

\acknowledgments

We are grateful for NASA funding from grant NNX10AJ56G and discretionary
founding from the McDonnell Center for the Space Sciences to build the
X-Calibur polarimeter. Q.Guo thanks the Chinese Scholarship Council from
China for the financial support (NO.2009629064) during her stay at
Washington University in St.\,Louis.
 

\bibliography{reportBeilicke}   
\bibliographystyle{spiebib}   

\end{document}